# From indexation policies through citation networks to normalized citation impacts: Web of Science, Scopus, and Dimensions as varying resonance chambers

Stephan Stahlschmidt[1] and Dimity Stephen[1*]

[1]German Centre for Higher Education Research and Science Studies (DZHW), Schützenstr. 6A, 10117, Berlin, Germany.

stahlschmidt@dzhw.eu, ORCID: 0000-0003-3390-8632
stephen@dzhw.eu, ORCID: 0000-0002-7787-6081, *Corresponding author

**Abstract**
Dimensions was introduced as an alternative bibliometric database to the well-established Web of Science (WoS) and Scopus, however all three databases have fundamental differences in coverage and content, resultant from their owners' indexation philosophies. In light of these differences, we explore here, using a citation network analysis and assessment of normalised citation impact of "duplicate" publications, whether the three databases offer structurally different perspectives of the bibliometric landscape or if they are essentially homogenous substitutes. Our citation network analysis of core and exclusive 2016-2018 publications revealed a large set of core publications indexed in all three databases that are highly self-referential. In comparison, each database selected a set of exclusive publications that appeared to hold similarly low levels of relevance to the core set and to one another, with slightly more internal communication between exclusive publications in Scopus and Dimensions than WoS. Our comparison of normalised citations for 41,848 publications indexed in all three databases found that German sectors were valuated as more impactful in Scopus and Dimensions compared to WoS, particularly for sectors with an applied research focus. We conclude that the databases do present structurally different perspectives, although Scopus and Dimensions with their additional circle of applied research vary more from the more base research-focused WoS than they do from one another.

**Declarations**

*Funding:* Data for WoS and Scopus in this study were obtained from the German Competence Centre for Bibliometrics (https://bibliometrie.info/), funded by the German Federal Ministry for Education and Research (BMBF) with grant number 01PQ17001.

*Competing interests*: The authors have no relevant interests to disclose.

*Availability of data, material and code*: Due to licensing agreements, the data in this study cannot be made publicly available.

*Authors' contributions*: Stephan Stahlschmidt conceived of and designed the study. Both authors collected the data, conducted the analyses, wrote and edited the manuscript, and read and approved the final manuscript.

*Ethics approval*: Not applicable

*Consent to participate*: Not applicable

*Consent for publication*: Not applicable

**Acknowledgements**
The authors would like to thank Digital Science for granting the DZHW access to raw Dimensions data.

# Introduction

Just as the introduction of Scopus in 2004 challenged the Web of Science's (WoS) position as the leading bibliometric database, the launch of Digital Science's Dimensions may change the bibliometric landscape once more. Like WoS and Scopus, Dimensions has amassed a huge index of scientific documents, however Dimensions has important fundamental differences to its predecessors that may offer a different bibliometric perspective. Further, Digital Science's position as part of the Holtzbrinck Publishing Group, owner of the SpringerNature publishing house, has embedded Dimensions in a content-rich environment not unlike Scopus' position within Elsevier. Given the potential uptake of Dimensions for bibliometric studies, it is important we understand how any diverging coverage of Dimensions, WoS, and Scopus might influence the results of bibliometric analyses.

Bibliometric analyses reflect their underlying databases' characteristics in that the databases' coverage fundamentally defines what is included in the analysis and bibliometric evaluation further contextualises the analysed content against the database, again emphasizing its coverage. As such, divergences between databases can have implications for bibliometric assessments. The primary differences between WoS, Scopus, and Dimensions for standard impact and productivity analyses lay in the scope of documents indexed as defined by the content selection processes. Consequently other influencing factors such as metadata accuracy, classification of document types, and discipline assignments are treated, in a statistical sense, as nuisance parameters that have to be controlled for in order to improve the validity of the emphasized coverage analysis.

Clarivate Analytics operates WoS under its founder Eugene Garfield's law of concentration arguing that the majority of significant scientific literature is published in only a small number of journals and only these journals must be indexed to achieve sufficient coverage of a discipline (Garfield, 1971; "Editorial selection process"). Clarivate's Emerging Source Citation Index and the diverse regional indices part from this concept without affecting the standard indices Science Citation Index Expanded, Social Science Citation Index and Arts & Humanities Index focused on in this study. Elsevier describes Scopus as the "most comprehensive overview of the world's research outputs", intending it to be the largest possible database of research items of acceptable quality (Elsevier, 2020), while Digital Science planned that Dimensions should be "open to integrate all relevant research objects", excluding predatory or otherwise unfit journals (Bode, Herzog, Hook & McGrath, 2019). In line with these philosophies, WoS and Scopus use selection panels to curate their content to that of a certain quality – with WoS enacting a higher threshold than Scopus ("Editorial selection process", Elsevier, 2020) – whereas Digital Science applies minimal editorial judgement to what is indexed in Dimensions, allowing the user to choose their own filters. This not only affects the visibility of single documents and their respective authors but also influences bibliometric impact analyses because all three calculate citation counts from the database-specific coverage of indexed items. As such citation counts are influenced by the database coverage, while an underlying Mathew effect in the science system may result in higher citation counts with larger collections.

Once content is selected, there are also differences between databases in how it is classified into document types and assigned to disciplines. All three databases use their own disciplinary classifications, however content in WoS and Scopus is assigned based on the discipline(s) of the publishing journal and both have been criticised both for the lack of transparency in how their classifications are applied and for inaccuracies in classification (Wang & Waltman, 2015). In contrast, Digital Science classifies each item using natural language processing and AI technology (Bode et al, 2019; Hook, Porter & Herzog, 2018), which eliminates the common issues associated with categorising content from multidisciplinary journals. These differences in classification practices can have important implications for the set of documents against which a publication is normalized and compared in the context of each database. Additionally the respective document type classification may also constitute an important interaction effect as different citation practices between types, such as between an often excluded journal editorial and a typically included journal article, affect bibliometric analyses. While differences in these classifications of research outputs make for interesting topics on their own, we focus our analysis on coverage differences and only strive to control these nuisance parameters.

As Dimensions was only recently launched, so far only a small number of studies have examined the differences in coverage between it, WoS, and Scopus that result from these differing practices and philosophies. Orduña-Malea & Delgado-López-Cózar (2018) compared samples of Library and Information Science documents, authors, and journals in Scopus and Dimensions and found that coverage in Dimensions exceeded that of Scopus, but Dimensions recorded fewer citations, potentially due to missing document links. Thelwall (2018) tested Dimensions' retrieval of nearly 90,000 food science articles with DOIs in Scopus, alongside a random sample of 10,000 articles from 2012. Over 90% of the food science articles were captured in Dimensions, and the citation counts in both databases were highly correlated (0.9-1.0), leading Thelwall to conclude that Scopus and Dimensions were interchangeable on coverage and citations. Harzing (2019) compared Dimensions, Scopus, and WoS on retrieval of her own publication corpus and six key Business and Economics journals. She found Dimensions and Scopus were approximately equal in both their coverage and citation counts, and both produced higher measures than in WoS.

Visser et al. (2020) compared WoS and Dimensions to Scopus, finding that Dimensions was the largest database, although there was substantial overlap in content between Scopus and both WoS (overlap of 17.7 million documents) and Dimensions (21.3 million). However, the share of Dimensions content not in Scopus was nearly double (40.9%, 14.8 million) that of WoS (22.7%, 5.2 million). Martín-Martín, Thelwall, Orduña-Malea, and Delgado López-Cózar (2020) also noted that, while Scopus and Dimensions offered twice as much exclusive content as WoS, the databases contained a high degree of overlapping content; 75-78% of their sample overlapped in pair-wise comparisons and 66% was present in all three databases. The largest content convergence occurred in the hard sciences, where 73-78% was in all three databases, with more divergent content in the medical sciences and engineering (65% in all three databases), social sciences (54%), and humanities (41%).

Motivated by these observations of coverage differences, we now address the implications for bibliometric analyses arising from these differences, focusing on impact analyses. Coverage differences between WoS and Scopus have already been shown to influence the outcome of bibliometric analyses. In a previous study, we found that German sectors with a focus on applied sciences had higher citation impact when assessed in Scopus, while those more oriented toward basic research fared better in WoS (Stahlschmidt & Stephen, 2019). Huang et al (2020) also noted that there were significant changes in the citation-based rankings of universities depending on which database was used to assess them. Here we analyse and compare the database-specific citation graph as a resonance chamber for German publications to observe how the different indexation policies affect the normalized citation analyses routinely applied in evaluation studies.

Our first step in this examination was to analysis the databases' citation networks. Previous studies show that the divergent coverage between the databases results in different subsets of publications indexed in only one, two or all three databases (Visser et al., 2020; Martín-Martín, et al., 2020). The power set of these subsets consists of the set of publications and citation links between them that are contained in all three databases, partial intersections of publications and related citations in two databases, and the residual sets of exclusive publications and related citations indexed in only one database. We investigated the role of these subsets for themselves and other subsets. With respect to citation analyses, role is understood as relevance and impact, which can be measured for higher aggregates by artificially reducing citations to their Mertonian ideal ("give credit, where credit is due"), by citations between and within these subsets. Citations between subsets represent how information flows and how the publications indexed exclusively in a database are embedded in this information flow. The role of these publications in the overarching citation network, which is unknown in its population, can be identified and the added value of each database as a reflection of previously unobserved scientific communication can be approximated.

Using the context provided by the citation network, we then assessed the differences in normalized citation impact of duplicate publications between the three databases. In bibliometrics, priority is given to normalised indicators that evaluate a publication in relation to its environment. Due to the different coverage of the databases, environment-specific differences arise in the evaluation of the same publication. We therefore analysed the same publications in the different databases and determined how each publication's valuation changed given the environment of the databases against which it was normalised. In doing so, the stratified German science system served as a coordinate system to measure and interpret differences. The varying evaluation of the same content can be used to illustrate the structural differences in the databases, which is informative for interpreting bibliometric analyses (Stahlschmidt & Stephen, 2019). Through these two analysis, we examined whether the databases offer a slightly varying but essentially homogeneous representation of the general citation network and are therefore substitutes, or whether the databases show structurally different bibliometric perspectives.

**Methods**

The data used for our analyses were sourced from the German Competence Centre of Bibliometrics' (KB) in-house versions of WoS and Scopus. Access to the Dimensions raw data was provided by Digital Science. For WoS, we used the established indices Science Citation Index Expanded, the Social Science Citation Index and the Arts and Humanities Citation Index. Scopus and Dimensions databases are not organised into indices and as such, we used the relevant documents from the entire database. Dimensions data were a snapshot of the database as of September 2019 and WoS and Scopus data were snapshots as of April 2019.

*Citation network analysis*
We analysed articles and reviews published in 2016-2018 and their citations to 2016 publications resulting in a citation network subset defined by a three year citation window between 2016 and 2018. We restricted our analysis to articles and reviews, however a known issue in Dimensions is that all documents in journals are assigned document type "article" (Visser et al., 2020). For instance, Dimensions holds more documents of type "article" in 2016-2018 than the intersection of core publications jointly indexed by WoS, Scopus and Dimensions does. However, most of these documents do not include any references. In comparison, only 1% of WoS and 4% of Scopus articles and reviews in 2016-2018 had no source references, and no WoS and only 9 Scopus articles had no references at all. To control this nuisance parameter in our coverage-focused analysis, we selected articles and reviews indexed in WoS and Scopus, and articles in Dimensions with at least one reference. Thus for Dimensions we separated the substantial scientific contributions that build on and highlight former contributions via references from other journal content to improve the validity of the database comparison. Still this requirement of at least one reference for Dimensions articles likely constitutes a lower bound rather than a solution to the currently diverging document type classification in Dimensions, but WoS and Scopus have also been observed to disagree occasionally on document types (Donner, 2017).

We joined the three databases WoS, Scopus, and Dimensions via an exact string matching procedure based on DOIs. DOIs uniquely identify publications and are therefore suitable for matching purposes. DOIs recorded in bibliometric databases have partially been observed to include errors (Zhu et al., 2019; Akbaritabar & Stahlschmidt, 2019), e.g. non-unique DOIs or misread string characters in optical character recognition processes. However these issues might arise randomly and might not adversely affect any structural differences between databases. Instead DOI matching has been observed to produce highly valid matching results (Fraser & Hobert, 2019) and less than 7%, 10%, respectively 1% of articles and reviews in 2016-2018 indexed in WoS, Scopus, and Dimensions were missing DOIs.

The citation network approach was implemented by the indicators (1) out/in-degree, as a network analytical perspective on the micro level of publications, which was translated to the level of the subsets previously described by representing the respective distributions, and (2), as far as possible the indicator internal coverage, as an aggregated value on the database level. With this approach, in addition to the expected increased coverage in Dimensions due to its larger size, i.e. the pure "more" of communication, communicative characteristics of the databases can also be partially quantified, which provides useful context for the normalised citation analysis.

*Normalised citation analysis*
We selected the German sectors as the level at which we assessed the effect of database choice on normalised citation impact. Hence, instead of evaluating several national science systems by a single bibliometric database, we evaluated different bibliometric databases by a single national science system, i.e. by the stratification of the German system into sectors.

We first identified all articles published in 2016 affiliated with German institutions that were indexed in all three databases, which we refer to as "duplicate" publications. We retrieved the WoS-Scopus duplicates, which were identified by comparing hash values on a subset of metadata strings between the two databases. We then extracted all German articles published in 2016 from Dimensions and matched their DOIs to the DOIs from the WoS-Scopus duplicates to identify the documents in all three databases. We validated the DOI-based matches by calculating the Jaro-Winkler distance on the title strings between the three versions.

To assess the effect of the database's environment on the normalised citations, we calculated for every German duplicate from each database the number of citations the article received between 2016 and 2018 (observed citations), and the average number of citations received in these three years by all articles published in 2016 that were allocated to the same discipline (expected citations). We then calculated the difference Δ in normalised citations between databases as:

$$\Delta\, norm.\, cit. = \frac{obtained\ citations_i^{(S1)}}{expected\ citations_i^{(S1)}} - \frac{obtained\ citations_i^{(S2)}}{expected\ citations_i^{(S2)}} \qquad (1)$$

where *i* is each German duplicate, and *s* is the source database.

We normalised each duplicate's citations against documents of the same type and discipline within each database. This necessitated that we exclude unclassified documents. We also mapped the database-specific discipline classification to the common OECD Fields of Science classification and retained only duplicates that were assigned to the same discipline in all databases to control for the diverse discipline assignment of publications as a nuisance parameter. We then aggregated the duplicates to sectors on a whole-counting basis.

In normalising the observed citations in each database against the expected citations in the same database, we achieved a database-specific valuation of each article indexed in all three databases. The content of the database was influential here as the inclusion or exclusion of particular articles in the corpus may influence both the citations received by the article and the average citations received by all articles in the discipline, affecting the ratio between the two observations. In examining the difference in normalised citation impact between databases, we can examine how the same content is valuated differently between databases due to the database's environment and hence infer the databases' latent characteristics based on the coordinate system supplied by the German science system.

**Results**

*Citation network analysis*
We show in Figure 1 the number of articles and reviews published in 2016-2018 in each database, and the intersection and exclusive content in each combination of databases as identified through the matching process. According to the inclusion criteria detailed above, Dimensions indexed the largest number of publications (>6.3 million), followed closely by Scopus (5.9 million), while WoS includes substantially fewer documents (4.8 million). The difference between Dimensions and Scopus may actually be smaller than reported here as the aforementioned imperfect lower bound we applied requiring Dimensions articles to have at least one reference might still allow other non-article documents to be included.

It seems the degrees of freedom to differentiate a database by its exclusively indexed publications are limited. Indeed, the majority of publications (4.3 million) were indexed in all three databases and this set of core publications is by far the largest intersection, constituting 67%, 71%, and 88% of the entire Dimensions, Scopus, and WoS corpuses, respectively. On top of the jointly indexed publications, Dimensions indexed an additional 1.3 million exclusive publications, or about 20% of its entire corpus, and Scopus exclusively indexed 0.6 million publications or approximately 10% of its corpus, while WoS' exclusive 37,566 publications constituted less than 1% of its corpus. Hence WoS differentiated itself from the other databases not by exclusively indexed publications, but seemingly by foregoing the indexation of more publications. Notably, contrary to its inclusive indexing policy, the Dimensions snapshot from April 2019 applied here apparently does not index some 1.05 million documents included in WoS and/or Scopus.

Given these sizable differences between the databases, we analysed how the internal coverage of core publications varied due to the different indexation practices. By observing if a reference in an indexed publication was or was not itself indexed in the same database, we compared the relevance attributed to the work by the author with the relevance attributed by the database provider. A pronounced difference in assumed relevance, manifesting as low coverage, indicates that a database only partially captures the communication flow perceived relevant by authors and hence a bibliometric analysis might be less informative on any such out-of-sync dataset.

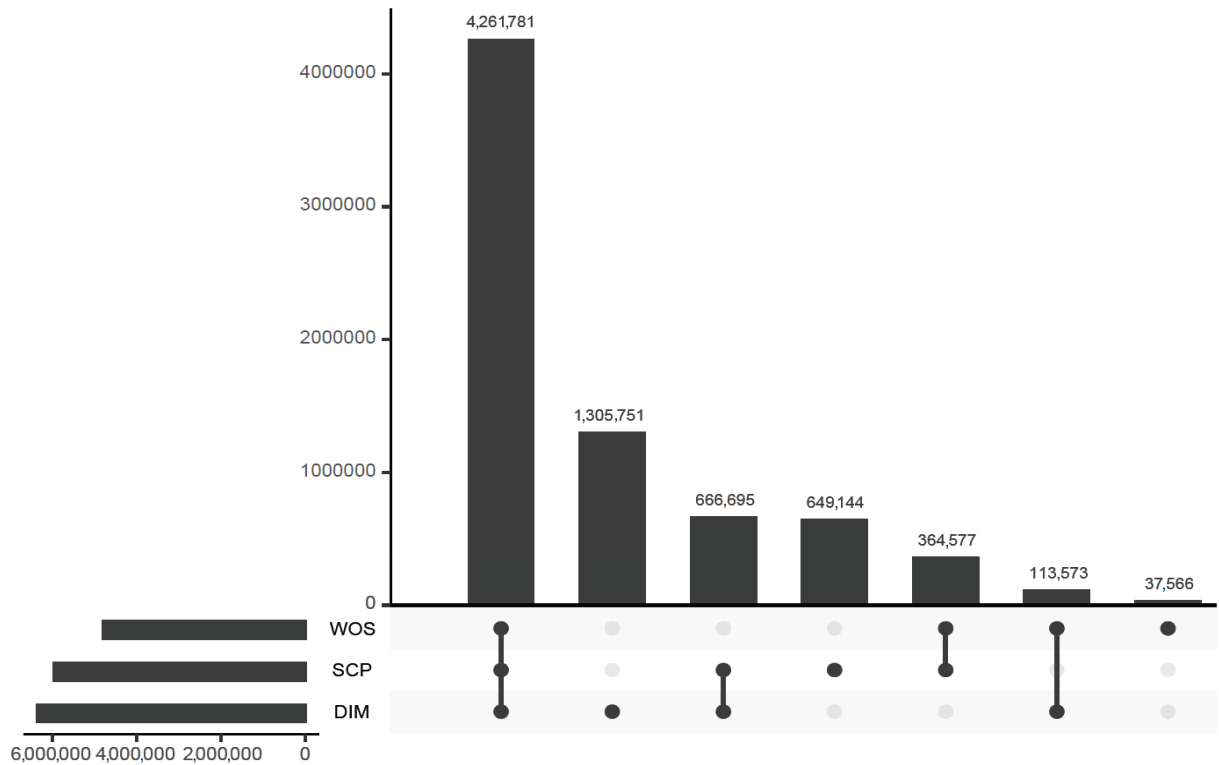

**Fig. 1** Number of publications with a DOI in 2016-2018 (left), magnitude of intersections (top) and relation between total databases and intersections (bottom)

As documents in the core set of jointly indexed publications have unanimously been deemed relevant by the three database providers and hence allow for a comparison, we used their reference lists to observe potential differences in the database-specific coverage. References to non-core publications especially differentiate the databases in this coverage. Figure 2 shows the database-specific percentage of indexed, or so-called source references, via a density plot across the 4.3 million core publications. The upper panel shows the overall internal coverage of 2016-2018 publications, while the middle panel restricts the references to publications published in 2004 or earlier, and the lower panel depicts the share of indexed references published in 2016.

The upper panel shows large variability in the overall internal coverage with values from zero to one hundred percent. All three distributions are skewed to the left, indicating that although a large share of references in core publications were indexed, some indexed core publications had few of their references indexed in the respective database. In particular, the social sciences and humanities, with their minor focus on journal articles as the primary communication device, have been observed to lack internal coverage (Kulczycki et al., 2018). Apart from this discipline-specific effect affecting all databases, we observed notable differences in the overall internal coverage between databases: whereas Dimensions and WoS demonstrated relatively high internal coverage, with over 90% of references from many core publications also indexed, Scopus exhibited lower agreement with the authors' relevance attribution.

This peculiarity of Scopus vanished when we restricted the set of references to those published in 2004 or earlier, as shown in the middle panel. Here, all three databases showed a similar, albeit lower, coverage of this reduced set of references. The previously described additional exclusive content in Scopus and Dimensions only slightly increased their internal coverage compared to WoS. However, the quality of Scopus data, a service launched in 2004 and covering publications back to at least 1996, has been observed to improve from publication year 2004 onwards (Stephen, Stahlschmidt & Hinze, 2020) and hence the lower coverage of Scopus in the upper panel might not result from cross-sectional differences in coverage, but rather from differences in coverage over time.

The lower panel shows the three-year citation window perspective as we restricted the analysis to references from 2016-2018 publications to 2016 publications. Again here we observed no pronounced visual difference between the databases, however, on average only approximately 5% of all references are considered and most other signals of relevance attribution via references are discarded in this typical three year citation window analysis. As the scale of the x-axis might conceal actual differences between the databases, in the following analyses we focused especially on this lower end of 2016 publications. In doing so we adopted a maximum contrast approach comparing core publications indexed in all three databases with exclusive publications indexed solely in one of the three databases.

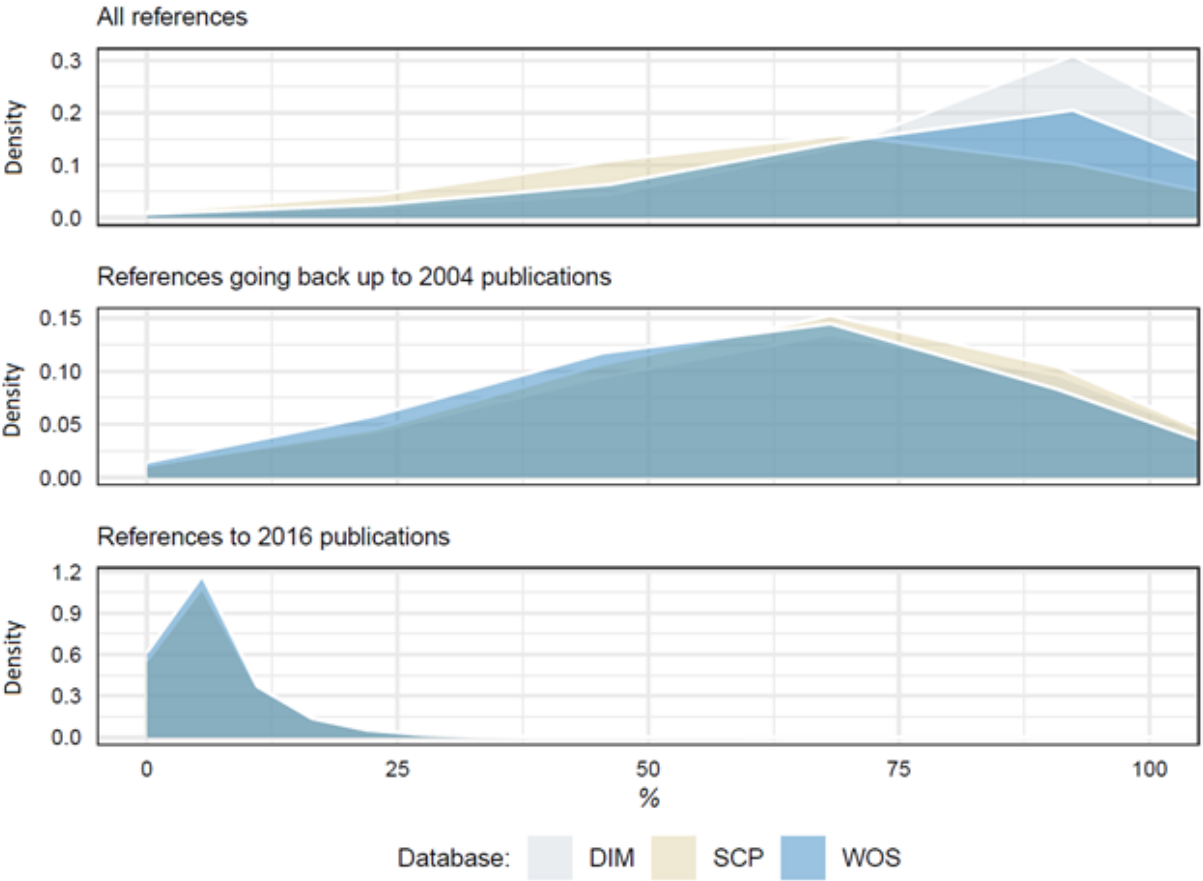

**Fig. 2** Internal coverage of references in jointly indexed publications published 2016-2018

We commenced the analysis of references to 2016 publications by examining the internal communication patterns within these subsets. Figure 3 shows the share of 2016-2018 core publications, jointly indexed in all three databases, that cited a core 2016 publication (top panel), and the share of 2016-2018 exclusive publications that cited an exclusive 2016 publication in the same database, highlighting the internal communication flow within these sets. As Digital Science currently only provides source references for publications indexed in Dimensions, the total number of references in exclusive Dimensions publications is unknown and for comparability we report instead the share of 2016 publications among all source references of 2016-2018 publications.

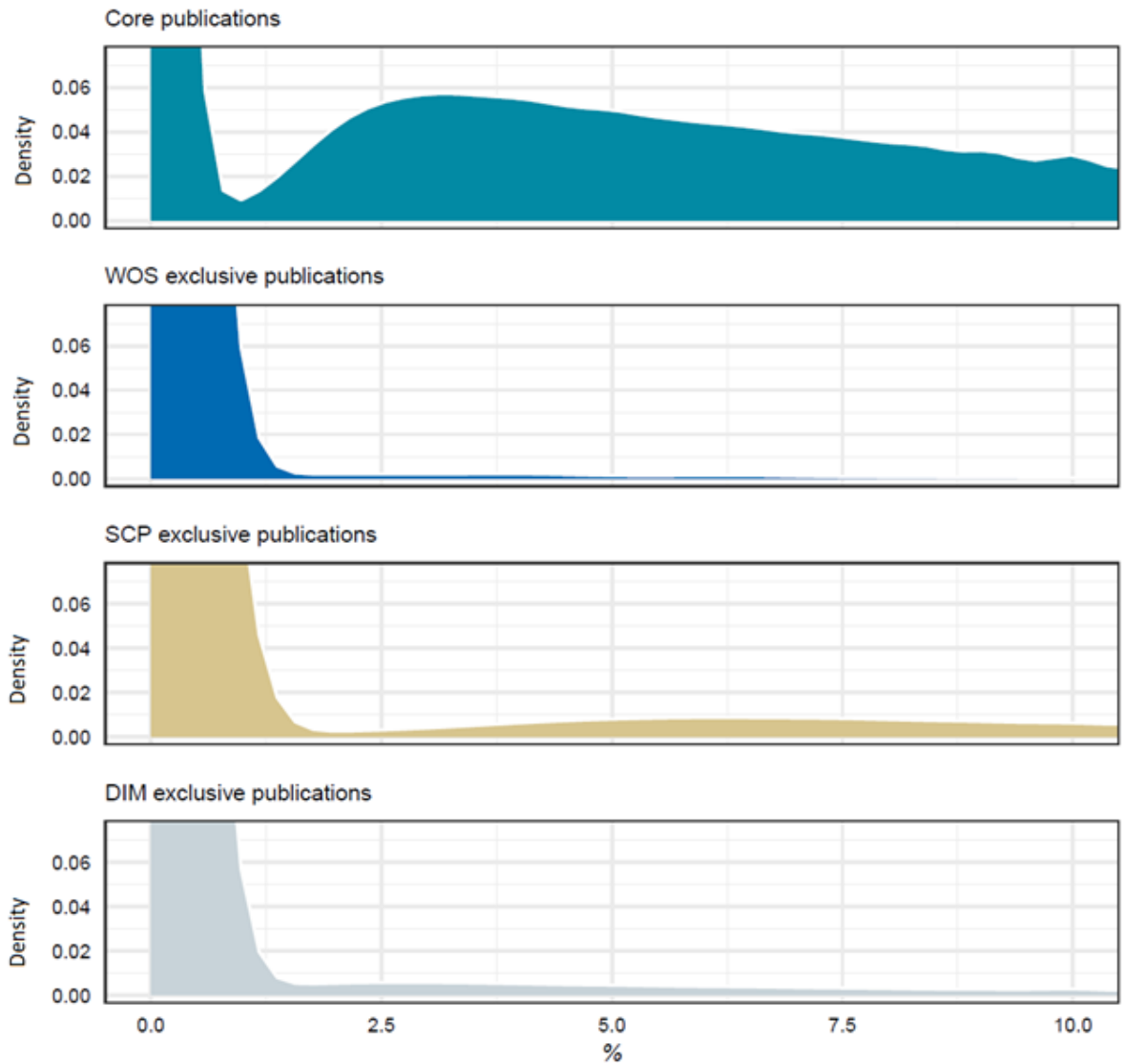

**Fig. 3** The share of internal references from core 2016-2018 publications to core 2016 publications (top) and from exclusive 2016-2018 publications to exclusive 2016 publications. Y-axis is truncated for visual clarity

As to be expected by the overall low share of references to 2016 publications depicted in the lower panel of Figure 2, the share of internal communication is rather small and the majority of 2016-2018 publications did not reference a single 2016 publication. This observation holds for all databases and hence WoS, Scopus and Dimensions can only be compared here and in the following analyses on the residual share of publications that did reference a 2016 publication. Comparing across the panels, we observed that core publications in the top panel possessed a much higher degree of internal communication than exclusive publications in the other panels, with a sizeable number of core publications referencing other core publications. Hence these publications define a highly self-referential, interlinked network component. This component was identified and indexed by all three databases and, as previously described, constituted between 67% (Dimensions) and 88% (WoS) of the databases.

To the contrary, publications exclusively indexed in one database differentiate the databases from one another, however in all databases these publications seldom referred to each other. Compared to core publications, the internal communication within these sets of database-specific exclusive publications was hardly visible, indicating they were only loosely connected by citation links, if at all. However, a slight difference in the internal communication in exclusive publications between databases is observable. WoS-exclusive publications were the least interconnected, while the much larger sets of Scopus- and Dimensions-exclusive publications communicated internally comparatively more often. In particular, the observed increase and local maximum around 6% in Scopus-exclusive publications indicates that Scopus indexed an additional component of the underlying citation network that is considerably more densely connected than the WoS or Dimensions equivalents. As such, Scopus has uniquely identified additional publications that show a relatively high degree of internal communication in parallel to the core publications and that might constitute a separate component of the underlying, but unidentified overall citation network.

Having assessed the internal communication of each set, we then examined how core and exclusive publications were interlinked. In Figure 4 we highlight the relevance of the core to exclusive publications by showing the share of references in exclusive 2016-2018 publications to core 2016 publications. As before, the majority of exclusive publications in 2016-2018 did not cite a 2016 core publication, so we had to discard most signals of relevance from the exclusive publications. However, we see that the share of references from exclusive to core publications is substantially higher than the share of internal communication within the exclusive publications depicted before in Figure 3. This observation holds for all three databases.

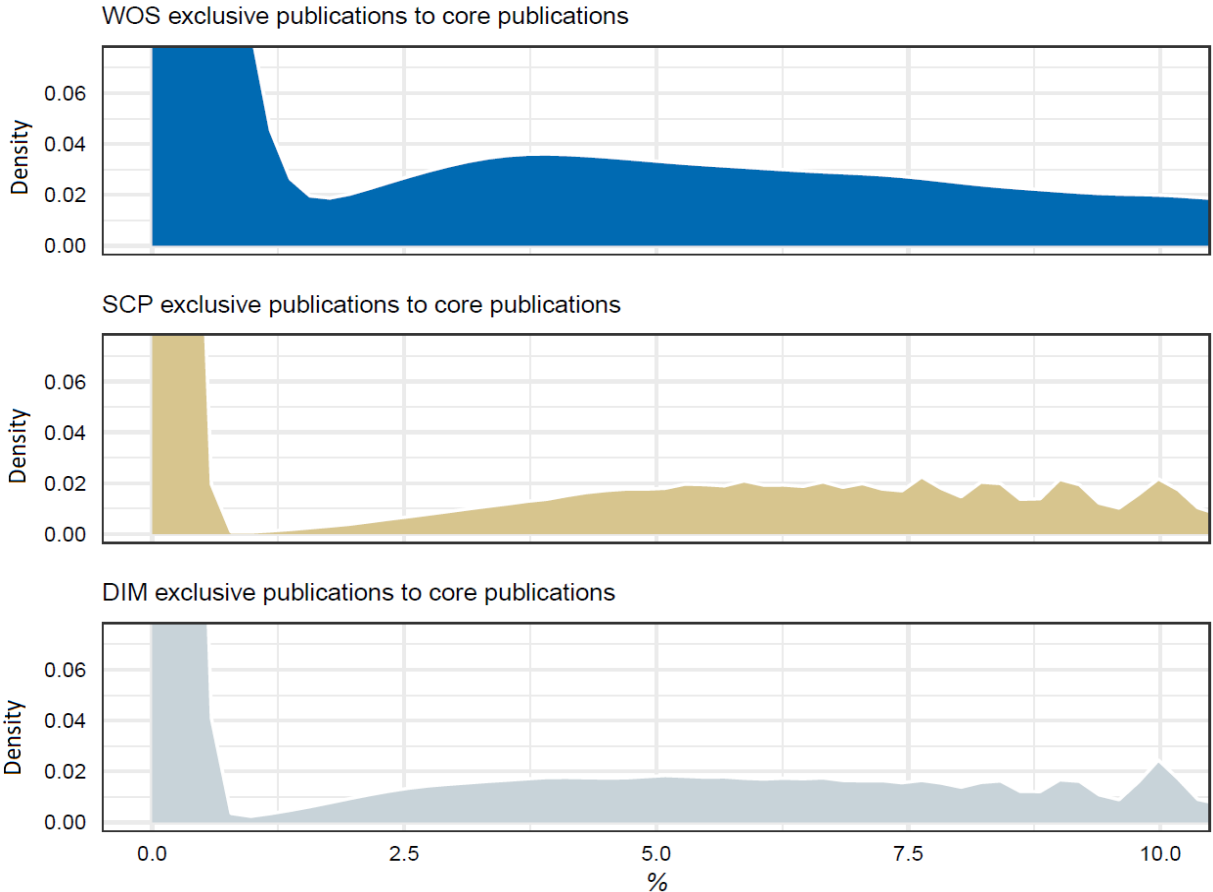

**Fig. 4** The share of references in exclusive 2016-2018 publications to core 2016 publications

Comparing the three databases, especially WoS-exclusive publications exhibited a strong dependence on core publications, resulting in a denser WoS citation graph than the Scopus and Dimensions equivalents. Although exclusive publications in these two larger databases also relied to a large and similar extent on core publications. Considering the different number of exclusive publications in each database (Figure 1), WoS seemingly foregoes indexing more publications, but offsets this with a more dense citation graph. Dimensions identified more exclusive publications than Scopus, which showed similar dependence on core publications as Scopus-exclusive publications (Figure 4) but slightly lower internal communication (Figure 3). Hence, the larger set of exclusive publications in Dimensions is offset by a sparser citation graph, while WoS holds a denser citation graph of substantially fewer publications.

To complete the interlinkage between core and exclusive publications, we depict in Figure 5 the share of references in core 2016-2018 publications citing exclusive 2016 publications. As we could observe the total number of references of these core publications via WoS and Scopus, we normalised by the total reference count and not the source reference count used before when the total number of references in the Dimensions-exclusive publications was unknown.

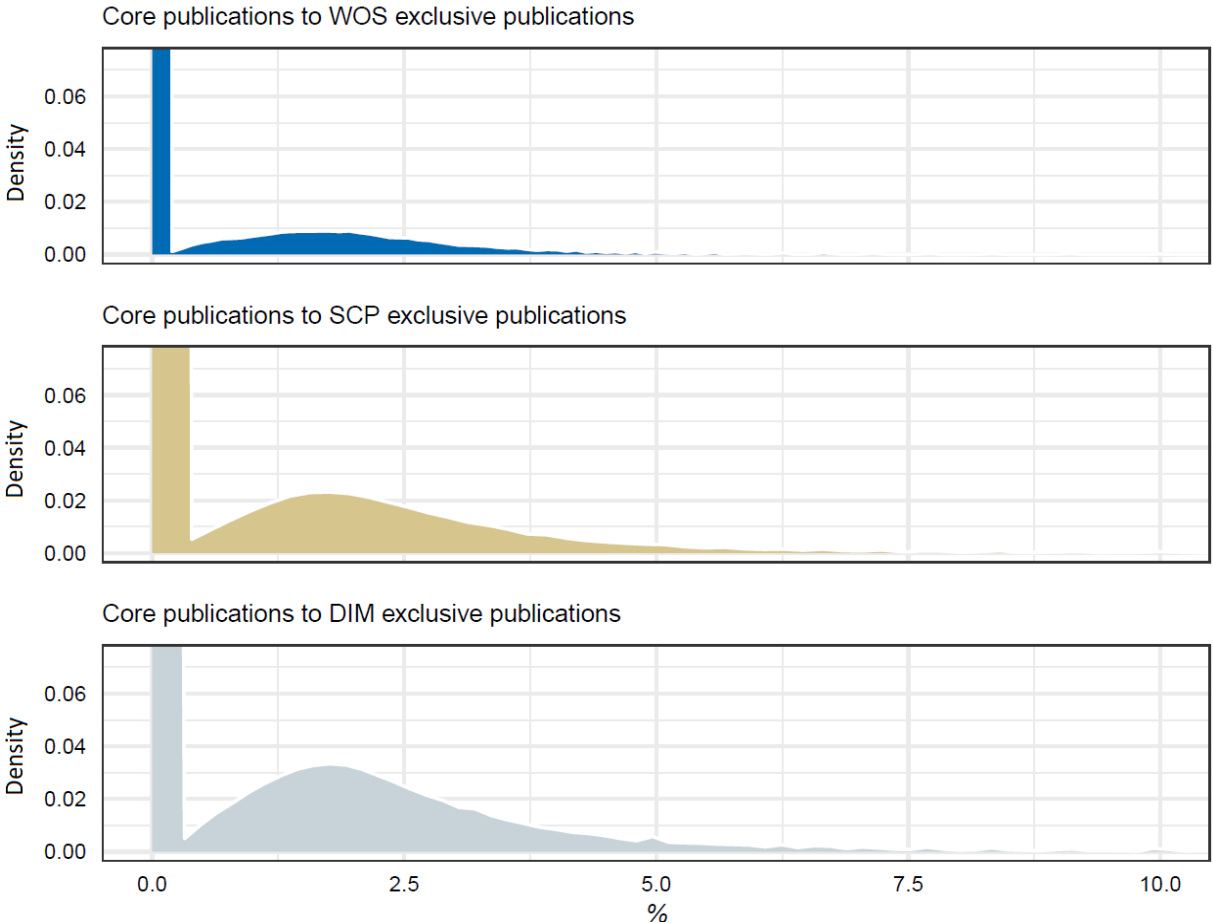

**Fig. 5** The share of references in core 2016-2018 publications to exclusive 2016 publications

A notable share of core references linked out to exclusive publications across all databases. However, considering the actual percentage is often less than 2.5% of all references, it appears that the exclusive publications are of low relevance to core publications. In comparison, the range for internal references among core publications in the upper panel of Figure 3 and the share of core publications referenced in exclusive publications in Figure 4 is much higher, often above 5%. As such, the content of core publications seemed to carry more relevance than the content of exclusive publications. Exclusive publications might then be understood as an outer, consecutive circle building upon the content provided in core publications while being of limited relevance to the core publications.

Although the databases covered a similar percentage range of references cited in core publications, the databases differed in the magnitude of the density in this range. The lower density in WoS may reflect that its 38,000 exclusive publications may not appear in the 4.3 million core publications as often as Dimensions' 1.3 million exclusive publications. Or, put differently, core publications referred to different exclusive publications, but in any case the share of these references was relatively low. Overall though, we observed a similar low relevance of exclusive publications in all databases and the particular choice of exclusive publications by a database resulted in no apparent difference to the core publications. That is, no database apparently managed to identify particularly relevant publications for the core publications, but instead all three databases found different exclusive publications of the same low relevance to the core. As a seeming paradox, exclusive publications in this sense did not distinguish databases from one another, but constituted different samples with the same characteristics. As a consequence, core publications might not be enhanced by any of the three perspectives offered by the databases.

*Normalised citation analysis*
For the citation analysis, we matched 2016 publications affiliated with a German institution across databases. The KB WoS-Scopus matching process identified 107,800 of the 113,227 in-scope WoS publications (95.2%) and 127,542 Scopus publications (84.5%) as duplicates. We identified 118,688 German publications in Dimensions, however this is an under-estimate of German publications as up to 50% of records in our Dimensions snapshot were missing information about the publishing country. We applied DOIs to match the WoS-Scopus duplicates to Dimensions. This process identified 84,332 publications that were in all 3 databases, which was 66.1% of the total 2016 German publications in Scopus, 71.1% in Dimensions, and 74.5% in WoS. After removing review documents, records missing discipline or sector data and documents assigned to different disciplines between databases, we had a final sample size of 41,848 duplicates.

To ensure the validity of the DOI matching, we calculated the Jaro-Winkler distance on the combinations of the three versions of the duplicates' titles. In all three comparisons, the distances between titles ranged from 0.0 to 0.73 with a mean of 0.02. We examined the 3,879 (4.6%) matches with distances above 0.25 and found that the higher distances resulted from different encoding of special characters, the inclusion of subtitles, or use of a German title in one database, but were otherwise correct matches. As such, we concluded that the DOI-based matching was accurate. It is possible that DOI matching falsely rejected some matching articles, however our sample size is sufficiently large that we expect our results to be robust to this likely small number of missing records.

As a preliminary examination of the differences in citations, we present in Figure 6 pair-wise comparisons of the citations observed for each duplicate article between databases. The diagonal lines represent a perfection correlation. Positive correlations were evident between each database, suggesting all three databases present a comparable picture of the bibliometric landscape, however there appeared to be a slight trend toward higher citations in Scopus and Dimensions compared to WoS. The greatest variation was between WoS and Scopus (mean difference = 1.1), with less between Dimensions and WoS (0.8), and least between Dimensions and Scopus (0.3). These variations from a linear trend due to the databases' exclusive content generate changes in the ratio of observed to expected citations, which translates to variations in the normalised citation impact of publications between databases.

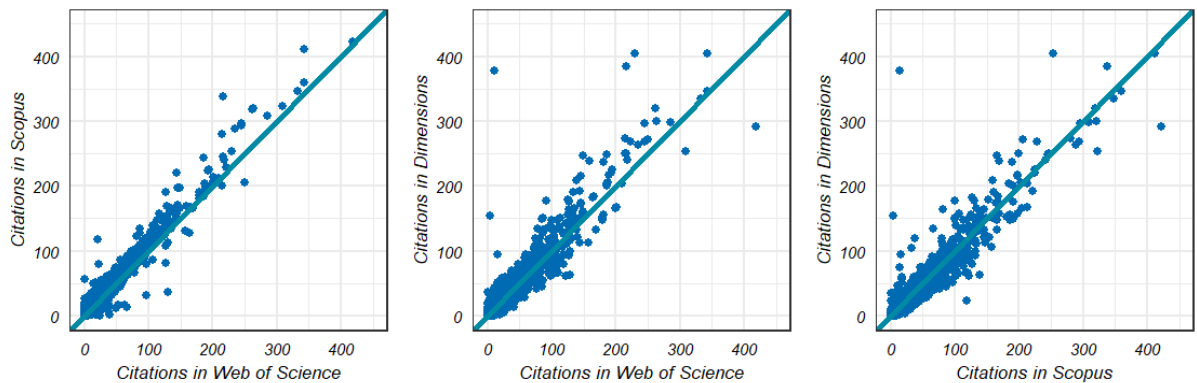

**Fig. 6** Pair-wise comparisons of duplicate articles' observed citations in each database

We see in Figure 7 the macro effect of the databases' structural differences via the differences in normalised citation impact detailed in formula (1) of the six sectors of the stratified German science system. Data are presented as the density in the distribution of differences in each sector's 2016 duplicate publications, and divided into quintiles.

The sectors are the Leibniz Association (WGL), the Max Planck Society (MPG), the Helmholtz Association (HGF), the Fraunhofer Society (FhG), universities and colleges of applied science constitute the higher education institutions (HEI), and the business sector (Economy). Each sector has a particular research profile. The universities undertake both teaching and research in all disciplines with unequivocal independence for the large number of small groups, respectively individual tenured staff, while the colleges focus on teaching and technical application in specific areas. The HGF has a health, energy, earth and physical sciences infrastructure orientation. The WGL includes a broad range of diverse and independent research institutes conducting research in all OECD FOS fields, while also providing research infrastructure and maintaining science museums. The MPG conducts primarily basic research, and the FhG focuses on applied research and transfer. Publications from the Economy group arise from private entities in their particular environment defined by specific business needs. Hence, we apply this stratification of the German science system as a coordinate system on base and applied research and according to their resonance with that system, pinpoint databases along these coordinates.

We see in the bottom panel of Figure 7 that the majority of all publications in each sector had higher normalised impact valuations in Scopus compared to WoS. The FhG and Economy sectors benefitted most strongly from Scopus's exclusive content, with around 70% of these sectors' publications improving in impact, compared to around 60% in the other sectors. Dimensions' content and characteristics similarly improved the impact of all sectors compared to WoS, as shown in the middle panel. The increased impact is nearly uniform across sectors, with the 40% of publications constituting the central and middle-high quintiles in each sector increasing in impact by up to 25%. Finally, the differences in normalised impact between Scopus and Dimensions in the top panel are more symmetrical, however there is a slight skew toward improved impact in Scopus, particularly for the FhG and MPG where nearly 60% of publications had increased impact.

Overall then, the larger content and specific characteristics of Scopus and Dimensions appears to produce higher normalised citations than in WoS, particularly for the sectors with a focus on applied sciences, such as the FhG and Economy sectors. However, the differences in content between Dimensions and Scopus produced less notable differences in normalised impact at this macro level.

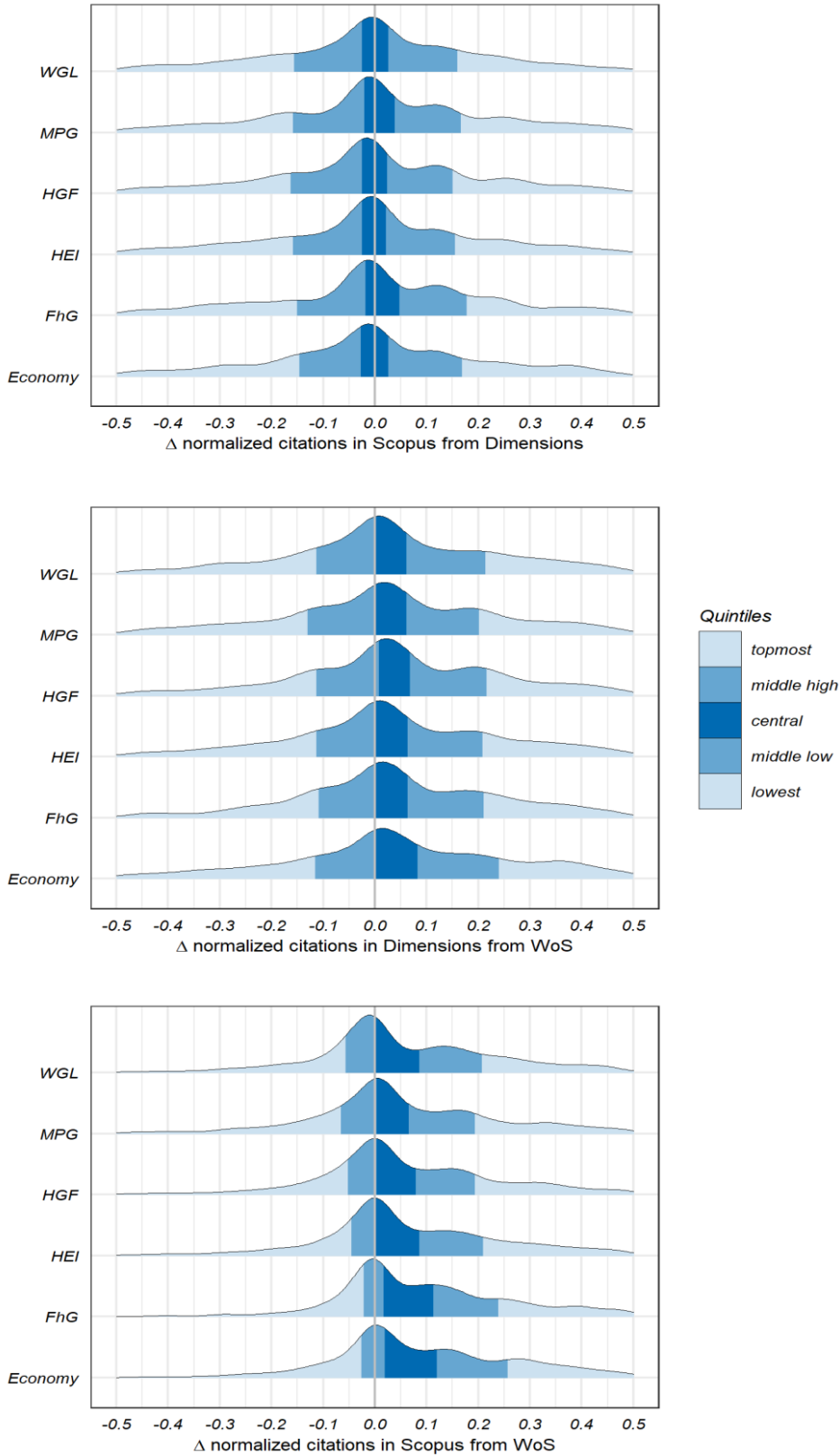

**Fig. 7** Distribution and quintile of differences in duplicates' normalised citations between databases by German sector

**Discussion**

WoS, Scopus, and Dimensions databases differ in particular fundamental characteristics, especially on the inclusion criteria applied to content and the resulting coverage, and the accuracy and classification of the accruing metadata. Each of these aspects is important as the environment of a database both defines how any particular publication is valued in terms of citations within the database, and also the context against which publications are normalised. Previous studies have found these differences resulted in variation in database size and coverage, although there was substantial overlap in the databases' content, and that the databases produced correlated citation counts (Martín-Martín et al., 2020; Visser et al., 2020; Orduña-Malea & Delgado-López-Cózar, 2018; Thelwall, 2018; Harzing, 2019). In this study, we analysed the subsequent consequences for the citation networks and normalised citation impact. We explored if the databases, via their exclusive content, offer structurally different or essentially homogeneous perspectives of the bibliometric landscape.

In our citation network analysis, we identified core publications jointly indexed in all three databases and exclusive publications solely indexed in one of the three databases. In a maximum contrast approach we compared the communication flows via references across these sets to observe the analytical potential each database revealed by its particular indexation practices. The indexation of more publications was accompanied by an offset in the density of communication flows in the resulting citation graph, particularly in the largest database Dimensions. Irrespective of the database however, we observed exclusive publications to mark an outer ring resting substantially upon core publications. Exclusive publications brought limited additional relevance to core publications, but the core might be characterised by its relatively high degree of self-reference, where former core publications constitute the base or frame for new core publications. Hence, knowledge is predominantly generated among the core publications and transferred from the core to the outer circle. This transfer was especially visible in Scopus and Dimensions. As the amount of parallel communication within exclusive publications seems negligible and the particular choice of exclusive publications did not affect its relation with the core, a star model seems to emerge of an interconnected core producing knowledge in a self-referential mode, which is then disseminated to the outer circle. The exact frontier between core publications and the outer ring might be difficult to define as a continuum between the two poles and several jump discontinuities in between arising from highly connected disciplines, geographic components or journal indexing practice impede an exact binary definition of each part. Scopus seems to adopt a slightly more extreme position on the continuum compared to the slightly fuzzier position of Dimensions, while the WoS coverage is focused almost exclusively on the core publications.

The combined effect of larger exclusive content in Scopus and Dimensions than WoS, similar reliance on core publications and hardly any relevance of exclusive to core publications increased the normalised citation impact of the same duplicated core publications of all German sectors in Scopus and Dimensions compared to WoS. However, in particular duplicated publications from German sectors with an applied focus benefited more strongly, highlighting the greater emphasis of Scopus and Dimensions on applied research among its non-core publications and testifying to the base research orientation of core publications defined especially by WoS. Scopus and Dimensions' similar levels of communication between and within the core and their choice of exclusive publications means there was little difference in impact between the databases given the respective indexation policies and subsequent coverage.

The methodology of our study intentionally highlighted the difference in coverage of the three databases and treats other influencing factors as controllable nuisance parameters. This somewhat dulled the effect of the databases' unique environments on the bibliometric indicators and motivates future studies on these factors, especially given the ongoing metadata quality improvements in Dimensions. Still our analysis partially explains why bibliometric databases report different impacts for the same diverse entities. For example the stable citation impact in WoS but yearly decreasing impact in Scopus of the base research focused MPG can be explained by our observations and exemplifies how the choice of a particular bibliometric database partially predetermines the outcomes of subsequent bibliometric analyses.

In this respect, WoS with its restrictive indexation policy and Scopus with its selective indexation policy constitute two separate self-imposed stances with a distinct message: WoS largely represents the well-interconnected core citation network component on base research, while Scopus allows us to observe some transfer from the core to the applied research periphery. Dimensions with its laissez-faire indexation policy conveys, apart from the improving metadata quality, more coverage but a similar, although less decisive, message to Scopus.

## References

Akbaritabar, A. & Stahlschmidt, S. (2019, September). Merits and limits: Applying open data to monitor Open Access publications in bibliometric databases. In G. Catalano, C. Daraio, M. Gregori, H. F. Moed, & G. Ruocco (Eds.), *Proceedings of the 17th International Conference on Scientometrics and Informetrics (Vol. 2)* (pp. 1455-1461). Rome: Edizioni Efesto.

Bode, C., Herzog, C., Hook, D. & McGrath, R. (2019). A guide to the Dimensions data approach: A collaborative approach to creating a modern infrastructure for data describing research: where we are and where we want to take it. Technical report. Digital Science. DOI: 10.6084/m9.figshare.5783094.

Bornmann. L. (2018). Field classification of publications in Dimensions: a first case study testing its reliability and validity. *Scientometrics 117*(1), 637–640. DOI:10.1007/s11192-018-2855-y.

Donner, P. (2017). Document type assignment accuracy in the journal citation index data of Web of Science. *Scientometrics*, 113(1), 219-236.

Editorial selection process. Retrived on March 5, 2020 from https://clarivate.com/webofsciencegroup/solutions/editorial/.

Elsevier. Scopus: Content Coverage Guide. Retrieved on July 13, 2020 from https://www.elsevier.com/__data/assets/pdf_file/0007/69451/Scopus_ContentCoverage_Guide_WEB.pdf.

Fraser N. & Hobert, A. (2019). Report on Matching of Unpaywall and Web of Science. Technical report. Kiel: ZBW

Garfield, E. (1971). The mystery of transposed journal lists – wherein Bradford's law of scattering is generalised according to Garfield's law of concentration. In: *Current Content 17*. Reprinted in Essays of an Information Scientist, pp. 222-223, by E. Garfield, 1977, Philadelphia: ISI Press.

Harzing, A. W. (2019). Two new kids on the block: How do Crossref and Dimensions compare with Google Scholar, Microsoft Academic, Scopus and the Web of Science? *Scientometrics, 120*(1), 341–349. DOI: 10.1007/s11192-019-03114-y.

Herzog, C. & Lunn, B. K. (2018). Response to the letter 'Field classification of publications in Dimensions: a first case study testing its reliability and validity'. *Scientometrics, 117*(1), 641–645. DOI: 0.1007/s11192-018-2854-z.

Hook, D., Porter, S. & Herzog, C. (2018). Dimensions: Building context for search and evaluation. *Frontiers in Research Metrics and Analytics, 3*(2). DOI: 10.3389/frma.2018.00023

Huang, C.-K., Neylon, C., Brookes-Kenworthy, C., Hosking, R., Montgomery, L., Wilson, K. & Ozaygen, A. (2020). Comparison of bibliographic data sources: Implications for the robustness of university rankings. *Quantitative Science Studies, 1*(2), 445–478. DOI: 10.1162/qss_a_00031.

Kulczycki, E., Engels, T.C.E., Pölönen, J., Bruun, K., Dušková, M., Guns, R., Nowotniak, R., Petr, M., Sivertsen, G., Istenič Starčič, A. and Zuccala, A. (2018) Publication patterns in the social sciences and humanities: evidence from eight European countries. *Scientometrics* 116**,** 463–486. https://doi.org/10.1007/s11192-018-2711-0

Martín-Martín, A., Thelwall, M., Orduna-Malea, E. & Delgado López-Cózar, E. (2020). Google Scholar, Microsoft Academic, Scopus, Dimensions, Web of Science, and OpenCitations' COCI: a multidisciplinary comparison of coverage via citations. *Scientometrics*, DOI: 10.1007/s11192-020-03690-4.

Orduña-Malea, E. & Delgado-López-Cózar, E. (2018). Dimensions: Re-discovering the ecosystem of scientific information. *El Profesional de la Información, 27*(2), 420-431. DOI: 10.3145/epi.2018.mar.21.

Stahlschmidt, S. & Stephen, D. (2019, September). Varying resonance chambers: A comparison of citation-based valuations of duplicated publications in Web of Science and Scopus. In G. Catalano, C. Daraio, M. Gregori, H. F. Moed, & G. Ruocco (Eds.), *Proceedings of the 17th International Conference on Scientometrics and Informetrics (Vol. 2)* (pp. 1698-1709). Rome: Edizioni Efesto.

Stephen, D., Stahlschmidt, S. & Hinze, S. (2020). Performance and Structures of the German Science System 2020. Studien zum deutschen Innovationssystem. Studie 5-2020. Berlin: Commission of Experts for Research and Innovation.

Thewall, M. (2018). Dimensions: A competitor to Scopus and the Web of Science? *Journal of Informetrics, 12*(2), 430-435. DOI: 10.1016/j.joi.2018.03.006.

Van Eck, N. J. & Waltman, L. (2017, September). Accuracy of citation data in Web of Science and Scopus. In R. Rousseau, W. Glänzel, & Z. Rongying (Eds.), *Proceedings of the 16th International Conference on Scientometrics and Informetrics* (pp. 1087-1092). Wuhan: HSE.

Visser, M., van Eck, N. J. & Waltman, L. (2020). *Large-scale comparison of bibliographic data sources: Scopus, Web of Science, Dimensions, Crossref, and Microsoft Academic*. Retrieved December 11, 2020 from: https://arxiv.org/abs/2005.10732.

Wang, Q. and Waltman, L. (2015). Large-scale analysis of the accuracy of the journal classification systems of Web of Science and Scopus. *Journal of Informetrics, 10*(2), 347–364. DOI: 10.1016/j.joi.2016.02.003

Zhu, J., Hu, G. & Liu, W. (2019). DOI errors and possible solutions for Web of Science. Scientometrics 118, 709–718. https://doi.org/10.1007/s11192-018-2980-7